\documentclass[letterpaper]{article}
\usepackage[margin=1.00in]{geometry}
\usepackage{mathtools}
\usepackage{amsmath,amsfonts,amssymb,amsthm,url,verbatim,cite}
\usepackage{verbatimbox}
\usepackage{graphicx,subcaption}
\usepackage{enumitem}
\usepackage{authblk}
\usepackage[ pdftex, plainpages = false, pdfpagelabels,
                 pdfpagelayout = useoutlines,
                 bookmarks,
                 bookmarksopen = true,
                 bookmarksnumbered = true,
                 breaklinks = true,
                 linktocpage=all,
                 pagebackref=false,
                 colorlinks = true,
                 linkcolor = BrickRed,
                 urlcolor  = blue,
                 citecolor = BrickRed,
                 anchorcolor = green,
                 hyperindex = true,
                 hyperfigures
                 ]{hyperref}
\usepackage[usenames, dvipsnames]{xcolor}
\usepackage{xifthen} 

\DeclarePairedDelimiter\ket{\lvert}{\rangle}

\DeclarePairedDelimiterX\braket[2]{\langle}{\rangle}{#1 \delimsize\vert #2}
\DeclarePairedDelimiterX\bbraket[2]{\langle\!\langle}{\rangle\!\rangle}{#1 \delimsize\| #2}
\DeclarePairedDelimiterX\cbraket[2]{(\!(}{)\!)}{#1 \delimsize\| #2}
\DeclarePairedDelimiterX\ketbra[2]{\lvert}{\rvert}{#1 \delimsize\rangle\!\langle #2}
\DeclarePairedDelimiterX\kketbra[2]{\|}{\|}{#1 \delimsize\rangle\!\rangle\!\langle\!\langle #2}
\DeclarePairedDelimiterX\cketbra[2]{\|}{\|}{#1 \delimsize)\!)\!(\!( #2}
\DeclarePairedDelimiterX\inner[2]{\langle}{\rangle}{#1,#2}

\def\tr{{\rm tr}\,}

\newcommand{\arxiv}[2][]{\ifthenelse{\isempty{#1}}{\href{http://arxiv.org/abs/#2}{{\tt arXiv:\allowbreak{}#2}}} {\href{http://arxiv.org/abs/#2}{{\tt arXiv:\allowbreak{}#2 [#1]}}}}

\newcommand{\booktitle}{\textsl}

\begin{document}
\title{Sporadic SICs and Exceptional Lie Algebras}
\author[$\dag$]{Blake C.\ Stacey}
\affil[$\dag$]{\small
  \href{http://www.physics.umb.edu/Research/QBism}{QBism Group},
  Physics Department, University of Massachusetts Boston, \par 100
  Morrissey Boulevard, Boston MA 02125, USA}
\date{\today}

\maketitle
\begin{abstract}
Sometimes, mathematical oddities crowd in upon one another, and the
exceptions to one classification scheme reveal themselves as
fellow-travelers with the exceptions to a quite different taxonomy.
\end{abstract}

\section{Preliminaries}

A set of \emph{equiangular lines} is a set of unit vectors in a
$d$-dimensional vector space such that the magnitude of the inner
product of any pair is constant:
\begin{equation}
  |\langle v_j, v_k\rangle| = \left\{\begin{array}{cc}
  1, & j = k; \\
  \alpha, & j \neq k.
  \end{array}\right.
\end{equation}
The maximum number of equiangular lines in a space of dimension $d$
(the so-called \emph{Gerzon bound}) is $d(d+1)/2$ for real vector
spaces and $d^2$ for complex. In the real case, the Gerzon bound is
only known to be attained in dimensions 2, 3, 7 and 23, and we know it
can't be attained in general. If you like peculiar alignments of
mathematical topics, the appearance of 7 and 23 might make your ears
prick up here. If you made the wild guess that the octonions and the
Leech lattice are just around the corner\ldots\ you'd be absolutely
right. Meanwhile, the complex case is of interest for quantum
information theory, because a set of $d^2$ equiangular lines in
$\mathbb{C}^d$ specifies a \emph{measurement} that can be performed
upon a quantum-mechanical system. These measurements are highly
symmetric, in that the lines which specify them are equiangular, and
they are ``informationally complete'' in a sense that quantum theory
makes precise. Thus, they are known as \emph{SICs}~\cite{Zauner:1999,
  Renes:2004, Scott:2010a, Fuchs:2017a}. Unlike the real case, where
we can only attain the Gerzon bound in a few sparse instances, it
\emph{appears} that a SIC exists for each dimension $d$, but nobody
knows for sure yet.

Before SICs became a physics problem, constructions of $d^2$ complex
equiangular lines were known for dimensions $d = 2$, 3 and 8. These
arose from topics like higher-dimensional polytopes and
generalizations thereof~\cite{Delsarte:1975, Hoggar:1981,
  Coxeter:1991, Hoggar:1998}.  Now, we have exact solutions for SICs
in the following dimensions~\cite{Appleby:2017, Appleby:2018,
  Kopp:2018, Grassl:2019}:
\begin{equation}
  \begin{array}{cl}
  d = &\!\!\!\!2\hbox{--}28, 30, 31, 35, \hbox{37--39}, 42, 43, 48, 49, 52,
  53, 57, 61\hbox{--}63, 67, 73, 74, 78, 79, 84, 91, 93, \\
  & \!\!95, \hbox{97--99},
  103, 109, 111, 120, 124, 127, 129, 134, 143, 146, 147, 168, 172,
  195, 199, \\
  & \!\!228, 259, 292, 323, 327, 399, 489, 844, 1299.
  \end{array}
\end{equation}
Moreover, numerical solutions to high precision are known for the
following cases:
\begin{equation}
  d = 2\hbox{--}189, 191, 192, 204, 224, 255, 288, 528, 725,
  1155, 2208.
\end{equation}
These lists have grown irregularly in the years since the
quantum-information community first recognized the significance of
SICs. (Many entries are due to A.\ J.\ Scott and
M.\ Grassl~\cite{Scott:2010a, Scott:2017, Grassl:2017}. Other pioneers
include M.\ Appleby, I.\ Bengtsson, T.-Y.\ Chien, S.\ T.\ Flammia,
G.\ S.\ Kopp and S.\ Waldron.) It is fair to say that researchers feel
that SICs \emph{should} exist for all integers $d \geq 2$, but we have
no proof one way or the other. The attempts to resolve this question
have extended into algebraic number theory~\cite{Kopp:2018,
  Appleby:2013, Appleby:2016, Bengtsson:2016, Appleby:2017b}, an
intensely theoretical avenue of research with the surprisingly
practical application of converting numerical solutions into exact
ones~\cite{Appleby:2017}. For additional (extensive) discussion, we
refer to the review article~\cite{Fuchs:2017a} and the
textbooks~\cite{Bengtsson:2017, Waldron:2018}.

In what follows, we will focus our attention mostly on the
\emph{sporadic SICs,} which comprise the SICs in dimensions 2 and 3,
as well as one set of them in dimension 8~\cite{Stacey:2016}. These
SICs have been designated ``sporadic'' because they stand out in
several ways, chiefly by residing outside the number-theoretic
patterns observed for the rest of the known
SICs~\cite{Appleby:2017b}. After laying down some preliminaries, we
will establish a connection between the sporadic SICs and the
exceptional Lie algebras $\mathrm{E}_6$, $\mathrm{E}_7$ and
$\mathrm{E}_8$ by way of their root systems.

\section{Quantum Measurements and Systems of Lines}

A \emph{positive-operator-valued measure} (POVM) is a set of
``effects'' (positive semidefinite operators satisfying $0 < E < I$)
that furnish a resolution of the identity:
\begin{equation}
\sum_i E_i = \sum_i w_i \rho_i = I,
\end{equation}
for some density operators $\{\rho_i\}$ and weights $\{w_i\}$. Note
that taking the trace of both sides gives a normalization constraint
for the weights in terms of the dimension of the Hilbert space. In
this context, the Born Rule says that when we perform the measurement
described by this POVM, we obtain the $i$-th outcome with probability
\begin{equation}
p(i) = \tr(\rho E_i),
\end{equation}
where $\rho$ without a subscript denotes our quantum state for the
system. The weighting $w_i$ is, up to a constant, the probability we
would assign to the $i$-th outcome if our state $\rho$ were the
maximally mixed state $\frac{1}{d}I$, the ``state of maximal
ignorance.''

SICs are a special type of POVM. Given a set of $d^2$ equiangular unit
vectors $\{\ket{\pi_i}\} \subset \mathbb{C}^d$, we can construct the
operators which project onto them, and in turn we can rescale those
projectors to form a set of effects:
\begin{equation}
  E_i = \frac{1}{d}\Pi_i, \hbox{ where } \Pi_i = \ketbra{\pi_i}{\pi_i}.
\end{equation}
The equiangularity condition on the $\{\ket{\pi_i}\}$ turns out to
imply that the $\{\Pi_i\}$ are linearly independent, and thus they
span the space of Hermitian operators on $\mathbb{C}^d$. Because the
SIC projectors $\{\Pi_i\}$ form a basis for the space of Hermitian
operators, we can express any quantum state $\rho$ in terms of its
(Hilbert--Schmidt) inner products with them. But, by the Born Rule,
the inner product $\tr(\rho \Pi_i)$ is, apart from a factor $1/d$,
just the probability of obtaining the $i$-th outcome of the SIC
measurement $\{E_i\}$.  The formula for reconstructing $\rho$ given
these probabilities is quite simple, thanks to the symmetry of the
projectors:
\begin{equation}
  \rho = \sum_i \left[(d+1)p(i) - \frac{1}{d}\right]\Pi_i,
\end{equation}
where $p(i) = \tr(\rho E_i)$ by the Born Rule. This furnishes us with
a map from quantum state space into the probability simplex, a map
that is one-to-one but not onto. In other words, we can fix a SIC as a
``reference measurement'' and then transform between density matrices
and probability distributions without ambiguity, but the set of
\emph{valid} probability distributions for our reference measurement
is a proper subset of the probability simplex.

We don't \emph{need} equiangularity for informational completeness,
just that the $d^2$ operators which form the reference measurement are
linearly independent and thus span the operator space. But
equiangularity implies the linear independence of those operators, and
it makes the formula for reconstructing $\rho$ from the overlaps
particularly clean~\cite{Appleby:2016b, DeBrota:2018}.

Because we can treat quantum states as probability distributions, we
can apply the concepts and methods of probability theory to them,
including Shannon's theory of information. The structures that I will
discuss in the following sections came to my attention thanks to
Shannon theory. In particular, the question of recurring interest is,
``Out of all the extremal states of quantum state space --- i.e., the
`pure' states $\rho = \ketbra{\psi}{\psi}$ --- which \emph{minimize}
the Shannon entropy of their probabilistic representation?'' I will
focus on the cases of dimensions 2, 3 and 8, where the so-called
sporadic SICs occur. In these cases, the information-theoretic
question of minimizing Shannon entropy leads to intricate geometrical
structures.

Any time we have a vector in $\mathbb{R}^3$ of length 1 or less, we
can map it to a $2 \times 2$ Hermitian matrix by the formula
\begin{equation}
  \rho = \frac{1}{2}\left(I + x\sigma_x + y\sigma_y + z\sigma_z\right),
\end{equation}
where $(x,y,z)$ are the Cartesian components of the vector and
$(\sigma_x, \sigma_y, \sigma_z)$ are the Pauli matrices. This yields a
positive semidefinite matrix $\rho$ with trace equal to 1; when the
vector has length 1, we have $\rho^2 = \rho$, and the density matrix
is a rank-1 projector that can be written as $\rho =
\ketbra{\psi}{\psi}$ for some vector $\ket{\psi}$.

Given any polyhedron of unit radius or less in $\mathbb{R}^3$, we can
feed its vertices into the Bloch representation and obtain a set of
density operators (which are pure states if they lie on the surface of
the Bloch sphere). For a simple example, we can do a regular
tetrahedron. Let $s$ and $s'$ take the values $\pm 1$, and define
\begin{equation}
  \rho_{s,s'} = \frac{1}{2}\left(I + \frac{1}{\sqrt{3}}
  (s\sigma_x + s'\sigma_y + ss' \sigma_z)\right).
\end{equation}
To make these density matrices into a POVM, scale them down by the
dimension. That is, take
\begin{equation}
E_{s,s'} = \frac{1}{2} \rho_{s,s'}.
\end{equation}
Then, the four operators $E_{s,s'}$ will sum to the identity. In fact,
they comprise a SIC.

By introducing a sign change, we can define another SIC,
\begin{equation}
  \tilde{\rho}_{s,s'} = \frac{1}{2}\left(I + \frac{1}{\sqrt{3}}
  (s\sigma_x + s'\sigma_y - ss' \sigma_z)\right).
\end{equation}
Each state in the original SIC is orthogonal to exactly one state in
the second. In the Bloch sphere representation, orthogonal states
correspond to \emph{antipodal} points, so taking the four points that
are antipodal to the vertices of our original tetrahedron forms a
second tetrahedron. Together, the states of the two SICs form a cube
inscribed in the Bloch sphere.

Here we have our first appearance of Shannon theory entering the
story. With respect to the original SIC, the states
$\{\ket{\tilde{\pi}_i}\}$ of the antipodal SIC all minimize the
Shannon entropy. The two interlocking tetrahedra are, entropically
speaking, dual structures.

\section{$\mathrm{E}_6$}
In what follows, I will refer to H.\ S.\ M.\ Coxeter's
\booktitle{Regular Complex Polytopes}~\cite{Coxeter:1991}. Coxeter
devotes a goodly portion of chapter 12 to the \emph{Hessian
  polyhedron}, which lives in $\mathbb{C}^3$ and has 27
vertices. These 27 vertices lie on nine diameters in sets of three
apiece. (In a real vector space, only two vertices of a convex
polyhedron can lie on a diameter.  But in a complex vector space,
where a diameter is a complex line through the center of the
polyhedron, we can have more~\cite{Coxeter:1974}.) He calls the
polyhedron ``Hessian'' because its nine diameters and twelve planes of
symmetry interlock in a particular way. Their incidences reproduce the
\emph{Hesse configuration}, a set of nine points on twelve lines such
that four lines pass through each point and three points lie on each
line.

Coxeter writes the 27 vertices of the Hessian polyhedron explicitly,
in the following way. First, let $\omega$ be a cube root of unity,
$\omega = e^{2\pi i / 3}$. Then, construct the complex vectors
\begin{equation}
  (0, \omega^\mu, -\omega^\nu),\ (-\omega^\nu, 0, \omega^\mu),
  \ (\omega^\mu, -\omega^\nu, 0),
\end{equation}
where $\mu$ and $\nu$ range over the values 0, 1 and 2. As Coxeter
notes, we could just as well let $\mu$ and $\nu$ range over 1, 2 and
3. He prefers this latter choice, because it invites a nice notation:
We can write the vectors above as
\begin{equation}
0\mu\nu,\ \nu0\mu,\ \mu\nu0.
\end{equation}
For example,
\begin{equation}
230 = (\omega^2, -1, 0),
\end{equation}
and
\begin{equation}
103 = (-\omega, 0, 1).
\end{equation}
Coxeter then points out that this notation was first introduced by
Beniamino Segre, ``as a notation for the 27 lines on a general cubic
surface in complex projective 3-space. In that notation, two of the
lines intersect if their symbols agree in just one place, but two of
the lines are skew if their symbols agree in two places or nowhere.''
Consequently, the 27 vertices of the Hessian polyhedron correspond to
the 27 lines on a cubic surface ``in such a way that two of the lines
are intersecting or skew according as the corresponding vertices are
non-adjacent or adjacent.''

Every smooth cubic surface in the complex projective space
$\mathbb{C}P^3$ has exactly 27 lines that can be drawn on it. For a
special case, the \emph{Clebsch surface}, we can actually get a real
surface (that is, in~$\mathbb{R}P^3$) that we can mold in plaster and
contemplate in the peaceful stillness of a good library. Intriguingly,
the coefficients for the lines on the Clebsch surface live in the
``golden field'' $\mathbb{Q}(\sqrt{5})$, which we will meet again
later in this article.

Casting the Hessian polyhedron into the real space $\mathbb{R}^6$, we
obtain the polytope known as $2_{21}$, which is related to
$\mathrm{E}_6$, since the Coxeter group of $2_{21}$ is the Weyl group
of $\mathrm{E}_6$. The Weyl group of $\mathrm{E}_6$ can also be
thought of as the Galois group of the 27 lines on a cubic surface.

We make the connection to symmetric quantum measurements by following
the trick that Coxeter uses in his Eq.~(12.39). We transition from the
space $\mathbb{C}^3$ to the complex projective plane by collecting the
27 vertices into equivalence classes, which we can write in
homogeneous coordinates as follows:
\begin{equation}
  \begin{array}{ccc}
    (0, 1, -1), & (-1, 0, 1), & (1, -1, 0) \\
    (0, 1, -\omega), & (-\omega, 0, 1), & (1, -\omega, 0) \\
    (0, 1, -\omega^2), & (-\omega^2, 0, 1), & (1, -\omega^2, 0)
    \end{array}
\end{equation}
Let $u$ and $v$ be any two of these vectors. We find that
\begin{equation}
|\langle u, u\rangle|^2 = 4
\end{equation}
when the vectors coincide, and
\begin{equation}
|\langle u, v\rangle|^2 = 1
\end{equation}
when $u$ and $v$ are distinct. We can normalize these vectors to be
quantum states on a three-dimensional Hilbert space by dividing each
vector by $\sqrt{2}$.

We have found a SIC for $d = 3$. When properly normalized, Coxeter's
vectors furnish a set of $d^2 = 9$ pure quantum states, such that the
magnitude squared of the inner product between any two distinct states
is $1/(d+1) = 1/4$.

Every known SIC has a group covariance property. Talking in terms of
projectors, a SIC is a set of $d^2$ rank-1 projectors $\{\Pi_j\}$ on a
$d$-dimensional Hilbert space that satisfy the Hilbert--Schmidt inner
product condition
\begin{equation}
\tr (\Pi_j \Pi_k) = \frac{d\delta_{jk} + 1}{d+1}.
\end{equation}
These form a POVM if we rescale them by $1/d$. In every known case, we
can compute all the projectors $\{\Pi_j\}$ by starting with one
projector, call it $\Pi_0$, and then taking the orbit of $\Pi_0$ under
the action of some group. The projector $\Pi_0$ is known as the
\emph{fiducial state}. (I don't know who picked the word ``fiducial'';
I think it was something Carl Caves decided on, way back.)

In all known cases but one, the group is the \emph{Weyl--Heisenberg
  group} in dimension $d$. To define this group, fix an orthonormal
basis $\{\ket{n}\}$ and define the operators $X$ and $Z$ such that
\begin{equation}
  X\ket{n} = \ket{n+1},
\end{equation}
interpreting addition modulo $d$, and
\begin{equation}
Z\ket{n} = e^{2\pi i n / d} \ket{n}.
\end{equation}
The Weyl--Heisenberg displacement operators are
\begin{equation}
D_{l\alpha} = (-e^{i\pi / d})^{l\alpha} X^l Z^\alpha.
\end{equation}
Because the product of two displacement operators is another
displacement operator, up to a phase factor, we can make them into a
group by inventing group elements that are displacement operators
multiplied by phase factors. This group has Weyl's name attached to
it, because he invented $X$ and $Z$ back in 1925, while trying to
figure out what the analogue of the canonical commutation relation
would be for quantum mechanics on finite-dimensional Hilbert
spaces~\cite{Weyl:1931, Scholz:2006, Scholz:2007}. It is also called
the \emph{generalized Pauli group}, because $X$ and $Z$ generalize the
Pauli matrices $\sigma_x$ and $\sigma_z$ to higher dimensions (at the
expense of no longer being Hermitian).

To relate this with the Coxeter construction we discussed earlier,
turn the first of Coxeter's vectors into a column vector:
\begin{equation}
\left(\begin{array}{c} 0 \\ 1 \\ -1\end{array}\right).
\end{equation}
Apply the $X$ operator twice in succession to get the other two
vectors in Coxeter's table (converted to column-vector format). Then,
apply $Z$ twice in succession to recover the right-hand column of
Coxeter's table. Finally, apply $X$ to these vectors again to effect
cyclic shifts and fill out the table. This set of nine states is known
as the \emph{Hesse SIC}.

Each of the 27 lines corresponds to a weight in the minimal
representation of~$\mathrm{E}_6$~\cite{Manivel:2006}, and so each
element in the Hesse SIC corresponds to three weights
of~$\mathrm{E}_6$.

In dimension $d = 3$, we encounter a veritable cat's cradle of
vectors~\cite{Stacey:2016c}.  First, there's the Hesse SIC.  Like all
informationally complete POVMs, it defines a probabilistic
representation of quantum state space, in this case mapping from $3
\times 3$ density matrices to the probability simplex for 9-outcome
experiments. As suggested earlier, we can look for the pure states
whose probabilistic representations minimize the Shannon entropy. The
result is a set of twelve states, which sort themselves into four
orthonormal bases of three states apiece. What's more, these bases are
\emph{mutually unbiased}: The Hilbert--Schmidt inner product of a
state from one basis with any state from another is always
constant. In a sense, the Hesse SIC has a ``dual'' structure, and that
dual is a set of Mutually Unbiased Bases (MUB).  This duality relation
is rather intricate: Each of the 9 SIC states is orthogonal to exactly
4 of the MUB states, and each of the MUB states is orthogonal to
exactly 3 SIC states~\cite{Stacey:2016c}.

An easy way to remember these relationships is to consider the finite
affine plane on nine points.

This configuration is also known as the discrete affine plane on nine
points, and as the Steiner triple system of order 3. That's a lot of
different names for something which is pretty easy to put together!
To construct it, first draw a $3 \times 3$ grid of points, and label
them consecutively:
\begin{equation}
  \begin{array}{ccc} 1 & 2 & 3 \\ 4 & 5 & 6 \\ 7 & 8 & 9 \end{array}
\end{equation}
These will be the points of our discrete geometry. To obtain the
lines, we read along the horizontals, the verticals and the leftward
and rightward diagonals:
\begin{equation}
\begin{array}{ccc} (123) & (456) & (789) \\ (147) & (258) & (369)
  \\ (159) & (267) & (348) \\ (168) & (249) & (357) \end{array}
\end{equation}
Each point lies on four lines, and every two lines intersect in
exactly one point. For our purposes today, each of the points
corresponds to a SIC vector, and each of the lines correponds to a MUB
vector, with point-line incidence implying orthogonality.  The four
bases are the four ways of carving up the plane into parallel lines
(horizontals, verticals, diagonals and other diagonals).

To construct a MUB vector, pick one of the 12 lines we constructed
above, and insert zeroes into those slots of a 9-entry probability
distribution, filling in the rest uniformly. For example, picking the
line $(123)$, we construct the probability distribution
\begin{equation}
  \left(0,0,0,\frac{1}{6},\frac{1}{6},\frac{1}{6},
  \frac{1}{6},\frac{1}{6},\frac{1}{6}
\right).
\end{equation}
This represents a pure quantum state that is orthogonal to the quantum state
\begin{equation}
  \left(\frac{1}{6},\frac{1}{6},\frac{1}{6},
  0,0,0,\frac{1}{6},\frac{1}{6},\frac{1}{6}
  \right)
  \end{equation}
and to
\begin{equation}
  \left(\frac{1}{6},\frac{1}{6},\frac{1}{6},
  \frac{1}{6},\frac{1}{6},\frac{1}{6},0,0,0 \right),
\end{equation}
while all three of these have the same Hilbert--Schmidt inner product
with the quantum state represented by
\begin{equation}
  \left(0,\frac{1}{6},\frac{1}{6},
  0,\frac{1}{6},\frac{1}{6},0,\frac{1}{6},\frac{1}{6} \right),
\end{equation}
for example.

Considering all the lines in the original structure that are
orthogonal to a given line in the dual yields a maximal set of real
equiangular lines in one fewer dimensions. (Oddly, I noticed this
happening up in dimension 8 before I thought to check in dimension
3~\cite{Stacey:2016b}, but we'll get to that soon.)  To visualize the
step from $\mathbb{C}^3$ to $\mathbb{R}^2$, we can use the Bloch
sphere representation for two-dimensional quantum state space. Pick a
state in the dual structure, i.e., one of the twelve MUB vectors. All
the SIC vectors that are orthogonal to it must crowd into a
2-dimensional subspace. In other words, they all fit into a
qubit-sized state space, and we can draw them on the Bloch
sphere. When we do so, they are coplanar and lie at equal intervals
around a great circle, a configuration sometimes called a
\emph{trine}~\cite{Caves:2004}. This configuration is a maximal
equiangular set of lines in the plane $\mathbb{R}^2$.

What happens if, starting with the Hesse SIC, you instead consider all
the lines in the dual structure that are orthogonal to a given vector
in the original?  This yields a SIC in dimension 2.  I don't know
where in the literature that is written, but it feels like something
Coxeter would have known.

Another path from the sporadic SICs to $\mathrm{E}_6$ starts with the
qubit SICs, i.e., regular tetrahedra inscribed in the Bloch sphere.
Shrinking a tetrahedron, pulling its vertices closer to the origin,
yields a type of quantum measurement (sometimes designated a
SIM~\cite{Graydon:2016}) that has more intrinsic noise. Apparently,
$\mathrm{E}_6$ is part of the story of what happens when the noise
level becomes maximal and the four outcomes of the measurement merge
into a single degenerate case. This corresponds to a singularity in
the space of all rotated and scaled tetrahedra centered at the
origin. Resolving this singularity turns out to involve the Dynkin
diagram of~$\mathrm{E}_6$: We invent a smooth manifold that maps to
the space of tetrahedra, by a mapping that is one-to-one and onto
everywhere except the origin. The pre-image of the origin in this
smooth manifold is a set of six spheres, and two spheres intersect if
and only if the corresponding vertices in the Dynkin diagram are
connected~\cite{Baez:2017}.

\section{$\mathrm{E}_8$} 

On the fourteenth of March, 2016, Maryna Viazovska published a proof
that the $\mathrm{E}_8$ lattice is the best way to pack hyperspheres
in eight dimensions~\cite{Viazovska:2017}. I celebrated the third
anniversary of this event by writing a guest post at the $n$-Category
Caf\'e, explaining how this relates to another packing problem that
seems quite different: how to fit as many equiangular lines as
possible into the \emph{complex} space $\mathbb{C}^8$. The answer to
this puzzle is another example of a SIC.

In the previous section, we saw how to build SICs by starting with a
fiducial vector and taking the orbit of that vector under the action
of a group, turning one line into $d^2$.  We said that the
Weyl--Heisenberg group was the group we use in call cases but
one. Now, we take on that exception.  It will lead us to the
exceptional root systems $\mathrm{E}_7$ and $\mathrm{E}_8$. Actually,
it will be a bit easier to tackle the latter first. Whence
$\mathrm{E}_8$ in the world of SICs?

We saw how to generate the Hesse SIC by taking the orbit of a fiducial
state under the action of the $d = 3$ Weyl--Heisenberg group. Next, we
will do something similar in $d = 8$. We start by defining the two
states
\begin{equation}
  \ket{\psi_0^\pm} \propto (-1 \pm 2i, 1, 1, 1, 1, 1, 1, 1)^{\mathrm{T}}.
\end{equation}

Here, we are taking the transpose to make our states column vectors,
and we are leaving out the dull part, in which we normalize the states
to satisfy
\begin{equation}
  \braket{\psi_0^+}{\psi_0^+} = \braket{\psi_0^-}{\psi_0^-} = 1.
\end{equation}

First, we focus on $\ket{\psi_0^+}$. To create a SIC from the fiducial
vector $\ket{\psi_0^+}$, we take the set of Pauli matrices, including
the identity as an honorary member: $\{ I, \sigma_x, \sigma_y,
\sigma_z \}$.  We turn this set of four elements into a set of
sixty-four elements by taking all tensor products of three
elements. This creates the Pauli operators on three qubits. By
computing the orbit of $\ket{\psi_0^+}$ under multiplication
(equivalently, the orbit of $\Pi_0^+ = \ketbra{\psi_0^+}{\psi_0^+}$
under conjugation), we find a set of 64 states that together form a
SIC set.

The same construction works for the other choice of sign,
$\ket{\psi_0^-}$, creating another SIC with the same symmetry
group. We can call both of them \emph{SICs of Hoggar type,} in honor
of Stuart Hoggar.

To make this connection, we consider the stabilizer of the fiducial
vector, i.e., the group of unitaries that map the SIC set to itself,
leaving the fiducial where it is and permuting the other $d^2 - 1$
vectors. Huangjun Zhu observed that the stabilizer of any fiducial for
a Hoggar-type SIC is isomorphic to the group of $3\times 3$ unitary
matrices over the finite field of order 9~\cite{Zhu:2012,
  Zhu:2015a}. This group is sometimes written $U_3(3)$ or
$\mathrm{PSU}(3,3)$. In turn, this group is up to a factor
$\mathbb{Z}_2$ isomorphic to $G_2(2)$, the automorphism group of the
Cayley integers, a subset of the octonions also known as the
\emph{octavians}~\cite{Conway:2003}. Up to an overall scaling, the
lattice of octavians is also the lattice known as $\mathrm{E}_8$.

The octavian lattice contains a great deal of arithmetic structure. Of
particular note is that it contains 240 elements of norm 1. In
addition to the familiar $+1$ and $-1$, which have order 1 and 2
respectively, there are 56 units of order 3, 56 units of order 6 and
126 units of order 4. The odd-order units generate subrings of the
octavians that are isomorphic to the \emph{Eisenstein integers} and
the \emph{Hurwitz integers,} lattices in the complex numbers and the
quaternions~\cite{Conway:2003}. From the symmetries of these lattices,
we can in fact read off the stabilizer groups for fiducials of the
qubit and Hesse SICs~\cite{Stacey:2016}. It is as if the sporadic SICs
are drawing their strength from the octonions.

Before moving on, we pause to note how peculiar it is that by trying
to find a nice packing of complex unit vectors, we ended up talking
about an optimal packing of Euclidean hyperspheres~\cite{Viazovska:2017}.

Now that we've met $\mathrm{E}_8$, it's time to visit the root system
we skipped: Where does $\mathrm{E}_7$ fit in?

\section{$\mathrm{E}_7$}

With respect to the probabilistic representation furnished by the
$\Pi_0^+$ SIC, the states of the $\Pi_0^-$ SIC minimize the Shannon
entropy, and vice versa~\cite{Stacey:2016b, Szymusiak:2015}.

\begin{myverbbox}{\vbfourlines}
1110111011100001111011101110000111101110111000010001000100011110
1101110111010010110111011101001011011101110100100010001000101101
1011101110110100101110111011010010111011101101000100010001001011
0111011101111000011101110111100001110111011110001000100010000111
\end{myverbbox}
\begin{table*}[t]
\begin{center}
\vbfourlines
\end{center}
\caption{\label{tab:4lines} Four of the states from the $\{\Pi_i^-\}$
  Hoggar-type SIC, written in the probabilistic representation of
  three-qubit state space provided by the $\{\Pi_i^+\}$ SIC. Up to an
  overall normalization by $1/36$, these states are all binary
  sequences, i.e., they are uniform over their supports.}
\end{table*}

Recall that when we invented SICs for a single qubit, they were
tetrahedra in the Bloch ball, and we could fit together two
tetrahedral SICs such that each vector in one SIC was orthogonal (in
the Bloch picture, antipodal) to exactly one vector in the other. The
two Hoggar-type SICs made from the fiducial states $\Pi_0^+$ and
$\Pi_0^-$ satisfy the grown-up version of this relation: Each state in
one is orthogonal to \emph{exactly twenty-eight states} of the other.

We can understand these orthogonalities as corresponding to the
antisymmetric elements of the three-qubit Pauli group. It is simplest
to see why when we look for those elements of the $\Pi_0^-$ SIC that
are orthogonal to the projector $\Pi_0^+$. These satisfy
\begin{equation}
\tr (\Pi_0^+ D \Pi_0^- D^\dagger) = 0
\end{equation}
for some operator $D$ that is the tensor product of three Pauli
matrices. For which such tensor-product operators will this expression
vanish? Intuitively speaking, the product $\Pi_0^+ \Pi_0^-$ is a
symmetric matrix, so if we want the trace to vanish, we ought to try
introducing an asymmetry, but if we introduce too much, it will cancel
out, on the ``minus times a minus is a plus'' principle. Recall the
Pauli matrices:
\begin{equation}
\sigma_x = \left(\begin{array}{cc} 0 & 1 \\ 1 &
  0 \end{array}\right),\qquad \sigma_y = \left(\begin{array}{cc} 0 &
  -i \\ i & 0 \end{array}\right),\qquad \sigma_z =
\left(\begin{array}{cc} 1 & 0 \\ 0 & -1 \end{array}\right).
\end{equation}
Note that of these three matrices, only $\sigma_y$ is antisymmetric,
and also note that we have
\begin{equation}
\sigma_z \sigma_x = -\sigma_x\sigma_z = i\sigma_y.
\end{equation}
This much is familiar, though that minus sign gets around. For
example, it is the fuel that makes the GHZ thought-experiment
go~\cite{Mermin:1993, Stacey:2018b}, because it means that
\begin{equation}
\sigma_x \otimes \sigma_x \otimes \sigma_x = -(\sigma_x \otimes
\sigma_z \otimes \sigma_z)(\sigma_z \otimes \sigma_x \otimes
\sigma_z)(\sigma_z \otimes \sigma_z \otimes \sigma_x).
\end{equation}
Let's consider the finite-dimensional Hilbert space made by composing
three qubits. This state space is eight-dimensional, and we build the
\emph{three-qubit Pauli group} by taking tensor products of the Pauli
matrices, considering the $2 \times 2$ identity matrix to be the
zeroth Pauli operator. There are 64 matrices in the three-qubit Pauli
group, and we can label them by six bits. The notation
\begin{equation}
\left(\begin{array}{ccc} m_1 & m_3 & m_5 \\ m_2 & m_4 & m_6\end{array}\right)
\end{equation}
means to take the tensor product
\begin{equation}
  \sigma_x^{m_1} \sigma_z^{m_2} \otimes (-i)^{m_3m_4} \sigma_x^{m_3}
  \sigma_z^{m_4} \otimes (-i)^{m_5m_6} \sigma_x^{m_5} \sigma_z^{m_6}.
\end{equation}
Now, we ask: Of these 64 matrices, how many are symmetric and how many
are antisymmetric? We can only get antisymmetry from $\sigma_y$, and
(speaking heuristically) if we include too much antisymmetry, it will
cancel out. More carefully put: We need an odd number of factors of
$\sigma_y$ in the tensor product to have the result be an
antisymmetric matrix. Otherwise, it will come out symmetric. Consider
the case where the first factor in the triple tensor product is
$\sigma_y$. Then we have $(4-1)^2 = 9$ possibilities for the other two
slots. The same holds true if we put the $\sigma_y$ in the second or
the third position. Finally, $\sigma_y \otimes \sigma_y \otimes
\sigma_y$ is antisymmetric, meaning that we have $9 \cdot 3 + 1 = 28$
antisymmetric matrices in the three-qubit Pauli group. In the notation
established above, they are the elements for which
\begin{equation}
m_1m_2 +m_3m_4 + m_5m_6 = 1 \mod 2.
\label{eq:antisymmetries}
\end{equation}
Moreover, these 28 antisymmetric matrices correspond exactly to the 28
bitangents of a quartic curve, and to pairs of opposite vertices of
the Gosset polytope $3_{21}$. In order to make this connection, we
need to dig into the octonions.

To recap: Each of the 64 vectors (or, equivalently, projectors) in the
Hoggar SIC is naturally labeled by a displacement operator, which up
to an overall phase is the tensor product of three Pauli operators.
Recall that we can write the Pauli operator $\sigma_y$ as the product
of~$\sigma_x$ and $\sigma_z$, up to a phase.  Therefore, we can label
each Hoggar-SIC vector by a pair of binary strings, each three bits in
length.  The bits indicate the power to which we raise the $\sigma_x$
and $\sigma_z$ generators on the respective qubits.  The pair
$(010,101)$, for example, means that on the three qubits, we act with
$\sigma_x$ on the second, and we act with $\sigma_z$ on the first and
third. Likewise, $(000, 111)$ stands for the displacement operator
which has a factor of~$\sigma_z$ on each qubit and no factors
of~$\sigma_x$ at all.

There is a natural mapping from pairs of this form to pairs of unit
octonions.  Simply turn each triplet of bits into an integer and pick
the corresponding unit from the set $\{1, e_1, e_2, e_3, e_4, e_5,
e_6, e_7\}$, where each of the $e_j$ square to $-1$.

We can choose the labeling of the unit imaginary octonions so that the
following nice property holds. Up to a sign, the product of two
imaginary unit octonions is a third, whose index is the \textsc{xor}
of the indices of the units being multiplied. For example, in binary,
$1 = 001$ and $4 = 100$; the \textsc{xor} of these is $101 = 5$, and
$e_1$ times $e_4$ is $e_5$.

Translate the Cayley--Graves table here into binary if enlightenment
has not yet struck:
\begin{equation}
\begin{array}{c|cccccccc}
e_i e_j & 1 & e_1 & e_2 & e_3 & e_4 & e_5 & e_6 & e_7 \\
\hline 
1 & 1 & e_1 & e_2 & e_3 & e_4 & e_5 & e_6 & e_7 \\
e_1 & e_1 & -1 & e_3 & -e_2 & e_5 & -e_4 & -e_7 & e_6 \\
e_2 & e_2 & -e_3 & -1 & e_1 & e_6 & e_7 & -e_4 & -e_5 \\
e_3 & e_3 & e_2 & -e_1 & -1 & e_7 & -e_6 & e_5 & -e_4 \\
e_4 & e_4 & -e_5 & -e_6 & -e_7 & -1 & e_1 & e_2 & e_3 \\
e_5 & e_5 & e_4 & -e_7 & e_6 & -e_1 & -1 & -e_3 & e_2 \\
e_6 & e_6 & e_7 & e_4 & -e_5 & -e_2 & e_3 & -1 & -e_1 \\
e_7 & e_7 & -e_6 & e_5 & e_4 & -e_3 & -e_2 & e_1 & -1
\end{array}
\end{equation}

So, each projector in the Hoggar SIC is labeled by a pair of
octonions, and the group structure of the displacement operators is,
almost, octonion multiplication. There are sign factors all over the
place, but for this purpose, we can neglect them. They will crop up
again soon, in a rather pretty way.

\begin{figure}[h]
  \begin{center}
    \includegraphics[width=8cm]{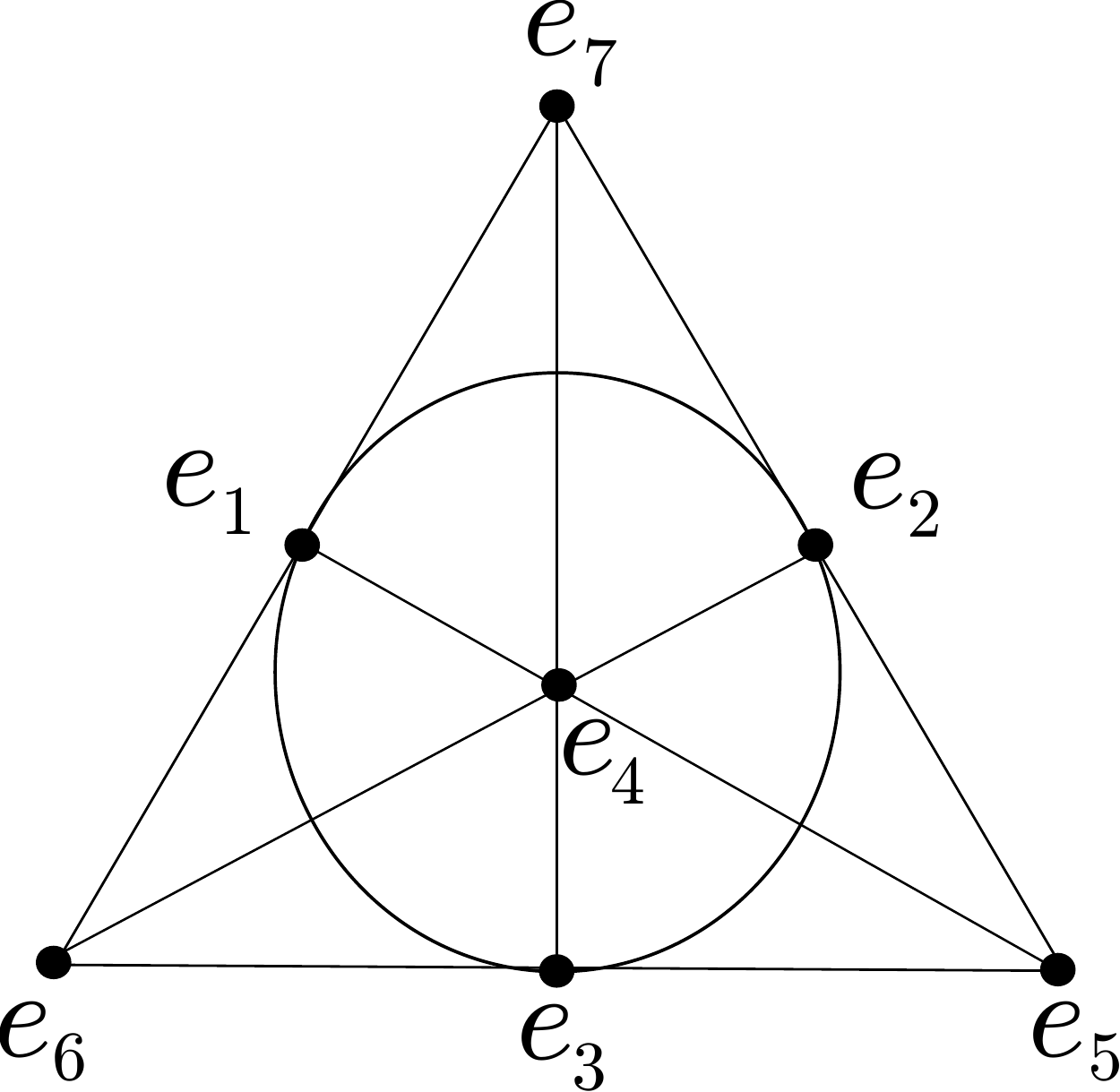}
  \end{center}
\caption{\label{fig:fano} The Fano plane: a symmetrical arrangement of
  seven points and seven lines.  Each point lies on three lines, each
  line contains three points, and every pair of lines intersect in a
  single point.}
\end{figure}

Another way to express the Cayley--Graves multiplication table is with
the \emph{Fano plane}, a set of seven points grouped into seven lines
that has been called ``the combinatorialist's coat of arms''. We can
label the seven points with the imaginary octonions $e_1$ through
$e_7$. When drawn on the page, a useful presentation of the Fano plane
has the point $e_4$ in the middle and, reading clockwise, the points
$e_1$, $e_7$, $e_2$, $e_5$, $e_3$ and $e_6$ around it in a regular
triangle. The three sides and three altitudes of this triangle, along
with the inscribed circle, provide the seven lines: $(e_1,e_2,e_3)$,
$(e_1,e_4,e_5)$, $(e_1,e_7,e_6)$, $(e_2,e_4,e_6)$, $(e_2,e_5,e_7)$,
$(e_3,e_4,e_7)$, $(e_3,e_6,e_5)$. It is apparent that each line
contains three points, and it is easy to check that each point lies
within three distinct lines, and that each pair of lines intersect at
a single point. One consequence of this is that if we take the
\emph{incidence matrix} of the Fano plane, writing a 1 in the $ij$-th
entry if line $i$ contains point $j$, then every two rows of the
matrix have exactly the same overlap:
\begin{equation}
  M = \begin{pmatrix}
     1 &1 &1 &0 &0 &0 &0 \\
     1 &0 &0 &1 &1 &0 &0 \\
     1 &0 &0 &0 &0 &1 &1 \\
     0 &1 &0 &1 &0 &1 &0 \\
     0 &1 &0 &0 &1 &0 &1 \\
     0 &0 &1 &1 &0 &0 &1 \\
     0 &0 &1 &0 &1 &1 &0
  \end{pmatrix}.
\end{equation}
The rows of the incidence matrix furnish us with seven equiangular
lines in~$\mathbb{R}^7$. We can build upon this by considering the
\emph{signs} in the Cayley--Graves multiplication table, which we can
represent by adding \emph{orientations} to the lines of the Fano
plane. Start by taking the first row of the incidence matrix $M$,
which corresponds to the line $(e_1, e_2, e_3)$, and give it all
possible choices of sign by multiplying by the elements not on that
line. Multiplying by $e_4$, $e_5$, $e_6$ and $e_7$ respectively, we
get
\begin{equation}
  \begin{pmatrix}
 + &+ &+ &0 &0 &0 &0 \\
 - &+ &- &0 &0 &0 &0 \\
 - &- &+ &0 &0 &0 &0 \\
 + &- &- &0 &0 &0 &0
  \end{pmatrix}.
\end{equation}
The sign we record here is simply the sign we find in the
corresponding entry of the Cayley--Graves table. Doing this with all
seven lines of the Fano plane, we obtain a set of 28 vectors, each one
given by a choice of a line and a point not on that line. Moreover,
\emph{all of these vectors are equiangular.} This is easily checked:
For any two vectors derived from the same Fano line, two of the terms
in the inner product will cancel, leaving an overlap of magnitude
1. And for any two vectors derived from different Fano lines, the
overlap always has magnitude 1 because each pair of lines always meets
at exactly one point. Van Lint and Seidel noted that the incidence
matrix of the Fano plane could be augmented into a full set of 28
equiangular lines~\cite{VanLint:1966, Seidel:1983}, but to my
knowledge, extracting the necessary choices of sign from octonion
multiplication is not reported in the literature.

So, the 28 lines in a maximal equiangular set in $\mathbb{R}^7$
correspond to point-line pairs in the Fano plane, where the point and
the line are \emph{not} coincident. In discrete geometry, the
combination of a line and a point not on that line is known as an
\emph{anti-flag.} It is straightforward to show from
Eq.~(\ref{eq:antisymmetries}) that the antisymmetric matrices in the
three-qubit Pauli group also correspond to the anti-flags of the Fano
plane: Simply take the powers of the $\sigma_x$ operators to specify a
point and the powers of the $\sigma_z$ operators to specify a
line~\cite{Stacey:2016b}.

Two fun things have happened here: First, we started with complex
equiangular lines. By carefully considering the orthogonalities
between two sets of complex equiangular lines, we arrived at a maximal
set of real equiangular lines in $\mathbb{R}^7$. And since one cannot
actually fit more equiangular lines into $\mathbb{R}^8$ than into
$\mathbb{R}^7$, we have a connection between a maximal set of
equiangular lines in $\mathbb{C}^8$ and a maximal set of them in
$\mathbb{R}^8$.

Second, our equiangular lines in $\mathbb{R}^7$ are the diameters of
the Gosset polytope $3_{21}$. And because we have made our way to the
polytope $3_{21}$, we have arrived at $\mathrm{E}_7$. To quote a
fascinating paper by Manivel~\cite{Manivel:2006},
\begin{quotation}
  \noindent Gosset seems to have been the first, at the very beginning
  of the 20th century, to understand that the lines on the cubic
  surface can be interpreted as the vertices of a polytope, whose
  symmetry group is precisely the automorphism group of the
  configuration. Coxeter extended this observation to the 28
  bitangents, and Todd to the 120 tritangent planes. Du Val and
  Coxeter provided systematic ways to construct the polytopes, which
  are denoted $n_{21}$ for $n = 2, 3, 4$ and live in $n + 4$
  dimensions. They have the characteristic property of being
  semiregular, which means that the automorphism group acts
  transitively on the vertices, and the faces are regular polytopes.
  In terms of Lie theory they are best understood as the polytopes in
  the weight lattices of the exceptional simple Lie algebras
  $\mathfrak{e}_{n+4}$, whose vertices are the weights of the minimal
  representations.
\end{quotation}

When we studied the Hesse SIC, we met the case $n = 2$ and
$\mathfrak{e}_6$. The intricate orthogonalities between two conjugate
SICs of Hoggar type have led us to the case $n = 3$ and $\mathfrak{e}_7$.

\section{The Regular Icosahedron and Real-Vector-Space Quantum Theory}

In the previous sections, we uncovered correspondences between
equiangular lines in $\mathbb{C}^3$ and $\mathbb{R}^2$, and between
$\mathbb{C}^8$ and $\mathbb{R}^7$. It would be nice to have a
connection like that between $\mathbb{C}^4$ and $\mathbb{R}^3$, but I
have not found one yet. Instead, there is a slightly different
relationship that brings $\mathbb{R}^3$ into the picture.

Suppose that, unaccountably, we wished to build the Hesse SIC, but in
\emph{real} vector space.  What might this even mean?  It would entail
finding a fiducial vector and an appropriate group, closely analogous
to the qutrit Weyl--Heisenberg group, such that the orbit of said
fiducial is a maximal set of equiangular lines.  How big would such a
set of lines be?  Recall that the Gerzon bound is $d^2$
for~$\mathbb{C}^d$, but only $d(d+1)/2$ in~$\mathbb{R}^d$.  In both
cases, this is essentially because those values are the dimensions of
the appropriate operator spaces (symmetric for operators
on~$\mathbb{R}^d$, self-adjoint for operators on~$\mathbb{C}^d$).  It
is not difficult to show that, if the Gerzon bound is attained, the
magnitude of the inner product between the vectors is $1/\sqrt{d+1}$
in~$\mathbb{C}^d$ and $1/\sqrt{d+2}$ in~$\mathbb{R}^d$.

We are familiar with the complex case, in which we define a shift
operator $X$ and a phase operator $Z$ that both have order $d$.  A
cyclic shift is nice and simple, so we'd like to keep that idea, but
the only ``phase'' we have to work with is the choice of positive or
negative sign.  So, let us consider the operators
\begin{equation}
  X = \begin{pmatrix} 0 & 0 & 1 \\ 1 & 0 & 0 \\ 0 & 1 &
    0 \end{pmatrix},
  \hbox{ and }
  Z = \begin{pmatrix} 1 & 0 & 0 \\ 0 & -1 & 0 \\ 0 & 0 &
    1 \end{pmatrix}.
\end{equation}
The shift operator $X$ still satisfies $X^3 = I$, while for the phase
operator $Z$, we now have $Z^2 = I$.

What group can we make from these operators?  Note that
\begin{equation}
  (ZX)^3 = -I,
\end{equation}
and so
\begin{equation}
  (-Z)^2 = X^3 = (-ZX)^3 = I,
\end{equation}
meaning that the operators $X$ and $-Z$ generate the \emph{tetrahedral
  group,} so designated because it is isomorphic to the rotational
symmetry group of a regular tetrahedron. Equivalently, we can use $Z$
as a generator, since $-Z = (ZX)^3 Z$ by the above.

Now, we want to take the orbit of a vector under this group!  But what
vector?  It should not be an eigenvector of~$X$ or of~$Z$, for then we
know we could never get a full set.  Therefore, we don't want a flat
vector, nor do we want any of the basis vectors, so we go for the next
simplest thing:
\begin{equation}
  v = \begin{pmatrix} 0 \\ 1 \\ y \end{pmatrix},
\end{equation}
where $y$ is a real number.  We now have
\begin{equation}
  Zv = \begin{pmatrix} 0 \\ -1 \\ y\end{pmatrix}
\end{equation}
and also
\begin{equation}
  X^2 v = \begin{pmatrix} 1 \\ y \\ 0 \end{pmatrix},
\end{equation}
so if we want equality between the inner products,
\begin{equation}
  \langle Zv, v \rangle = \langle{X^2 v, v \rangle},
\end{equation}
then we need to have
\begin{equation}
  -1 + y^2 = y.
\end{equation}
The positive solution to this quadratic equation is
\begin{equation}
  y = \frac{1 + \sqrt{5}}{2},
\end{equation}
so we can in fact take our $y$ to be $\phi$, the golden ratio.

In the group we defined above, $X$ performs cyclic shifts, $Z$ changes
the relative phase of the components, and we have the freedom to flip
all the signs.  Therefore, the orbit of the fiducial $v$ is the set of
twelve vectors
\begin{equation}
  \begin{pmatrix} 0 \\ \pm 1 \\ \pm \phi \end{pmatrix},
  \   \begin{pmatrix} \pm 1 \\ \pm \phi \\ 0 \end{pmatrix},
  \   \begin{pmatrix} \pm \phi \\ 0 \\ \pm 1 \end{pmatrix}.
\end{equation}
These are the vertices of a regular icosahedron, and the diagonals of
that icosahedron are six equiangular lines.  The inner products
between these vectors are always $\pm \phi$. Since $d(d+1)/2 = 6$,
there cannot be any larger set of equiangular lines in~$\mathbb{R}^3$.

Recall that the reciprocal of the golden ratio $\phi$ is
\begin{equation}
  \phi^{-1}
  = \frac{2}{1 + \sqrt{5}}
    \frac{1 - \sqrt{5}}{1 - \sqrt{5}}
    = \frac{2 - 2\sqrt{5}}{1 - 5}
    = \frac{-1 + \sqrt{5}}{2} = \phi - 1.
\end{equation}
The golden ratio $\phi$ is a root of the monic
polynomial $y^2 - y - 1$, and being a root of a monic polynomial with
integer coefficients, it is consequently an algebraic integer.  The
same holds for its reciprocal, so $\phi^{-1}$ is also an algebraic
integer, making the two of them \emph{units} in the number field
$\mathbb{Q}(\sqrt{5})$.

To summarize: For the diagonals of the regular icosahedron, the vector
components are given by the units of the ``golden field''
$\mathbb{Q}(\sqrt{5})$.  But it has been
discovered~\cite{Bengtsson:2016} that the vector components for the
Weyl--Heisenberg SICs in dimension \emph{four} are derived from a unit
in the \emph{ray class field over} $\mathbb{Q}(\sqrt{5})$.  Therefore,
in a suitably perplexing way, the icosahedron is the Euclidean version
of the Hesse SIC, and the SICs in $d = 4$ are the number-theoretic
extension of the icosahedron.

Futhermore, it has been observed empirically that in dimensions
\begin{equation}
  d_k = \phi^{2k} + \phi^{-2k} + 1,
\end{equation}
there exist Weyl--Heisenberg SICs with additional group-theoretic
properties that make their exact expressions easier to find. These are
known as \emph{Fibonacci--Lucas SICs}~\cite{Grassl:2017}.

There are exactly four known cases where the Gerzon bound can be
attained in~$\mathbb{R}^d$: when $d = 2$, 3, 7 and 23. Three out of
these four examples relate to SICs, specifically to the sporadic
SICs. We can obtain the maximal equiangular sets in~$\mathbb{R}^2$ and
$\mathbb{R}^7$ from SICs in~$\mathbb{C}^3$ and $\mathbb{C}^8$
respectively, while the set in~$\mathbb{R}^3$ turns out to be the real
analogue of our example in~$\mathbb{C}^3$. All of this raises a
natural question: What about~$\mathbb{R}^{23}$? Does the equiangular
set there descend from a SIC in~$\mathbb{C}^{24}$? That, nobody knows.

We do know that the maximal equiangular line set in~$\mathbb{R}^{23}$
can be extracted from the Leech lattice~\cite{Cohn:2007}. It contains
276 lines, and its automorphism group is Conway's group
$\mathrm{Co}_3$~\cite{Goethals:1975}.  Further study of this structure
connects back with our use of SICs to give a probabilistic
representation of quantum state space.  When we fix a SIC
in~$\mathbb{C}^d$ as a reference measurement, the condition $\tr\rho^2
= 1$, which is satisfied when $\rho$ is a pure state, becomes
\begin{equation}
\sum_i p(i)^2 = \frac{2}{d(d+1)}.
\end{equation}
Now, flipping this equation upside down makes both sides \emph{look
  like counting!} The right-hand side becomes combinatorics: It's just
the binomial coefficient for choosing two things out
of~$d+1$. Meanwhile, the left-hand side becomes the \emph{effective
  number} of outcomes, which we are familiar with because it is a
biodiversity index~\cite{Leinster:2011}:
\begin{equation}
  N_{\rm eff} = \left(\sum_i p(i)^2 \right)^{-1} .
\end{equation}
So, when we ascribe a pure state to a quantum system of Hilbert-space
dimensionality $d$, we are saying that the effective number of
possible outcomes for a reference measurement is
\begin{equation}
N_{\rm eff} = \binom{d+1}{2}.
\end{equation}
Consequently, ascribing a pure state means that we are effectively
ruling out a number of outcomes equal to
\begin{equation}
d^2 - N_{\rm eff} = \frac{d(d-1)}{2} = \binom{d}{2}.
\end{equation}
This motivates the following question: What is the upper bound on the
number of entries in $\vec{p}$ that can equal zero? A brief
calculation with the Cauchy--Schwarz inequality~\cite{Stacey:2016c}
reveals that the answer is, in fact, exactly $d^2 - N_{\rm eff}$. In
other words, no vector $\vec{p}$ in the image of quantum state space
can contain more than $d(d-1)/2$ zeros. One reason why the sporadic
SICs are distinguished from all the others is that they provide the
examples where this bound is known to be saturated. We can attain it
for qubit SICs (where it equals 1), the Hesse SIC (where it equals 3)
and the Hoggar-type SICs (where it equals 28). The states which
saturate this bound are also those that minimize the Shannon entropy,
as we discussed above.

We can also deduce the corresponding bound for the case of real
equiangular lines. In~$\mathbb{R}^d$, the Gerzon bound is $d(d+1)/2$,
and in those cases where we have a complete set of equiangular lines,
we can play the game of doing ``real-vector-space quantum mechanics'',
using our $d(d+1)/2$ equiangular lines to define a reference
measurement. As in~$\mathbb{C}^d$, the Cauchy--Schwarz inequality
gives an upper bound on the number of zero-valued entries in a
probability distribution $p$, which works out to be
\begin{equation}
N_Z = \frac{d^2 - 1}{3}.
\end{equation}
Thinking about it for a moment, we realize that this is telling us
about the maximum number of equiangular lines that can all
simultaneously be orthogonal to a common vector. In particular, if we
fix $d = 23$, we have a ``real-vector-space SIC'' that we can derive
from the Leech lattice, and we know that
\begin{equation}
N_Z = \frac{23^2 - 1}{3} = 176
\end{equation}
of the elements of that set can be orthogonal to a ``pure quantum
state'', i.e., a vector in~$\mathbb{R}^{23}$. All 176 such lines have
to crowd together into a 22-dimensional subspace, while still being
equiangular. They comprise a maximal set of equiangular lines
in~$\mathbb{R}^{22}$, whose symmetries form the Higman--Sims finite
simple group~\cite{Lemmens:1973}.

And that's what the classification of finite simple groups has to do
with biodiversity!

\section{Open Puzzles Concerning Exceptional Objects}
While we're thinking about equiangular lines in spaces other than
$\mathbb{C}^d$, here is a puzzle: What about the octonionic space
$\mathbb{O}^3$, which figures largely in the study of exceptional
objects? The Gerzon bound for this space works out to be 27. Cohn,
Kumar and Minton give a nonconstructive proof that a set saturating
the Gerzon bound in $\mathbb{O}^3$ exists, along with a numerical
solution~\cite{Cohn:2016}, but that numerical solution doesn't look
like an approximation of a really pretty exact solution in any obvious
way. (Their set of mutually unbiased bases in $\mathbb{O}^3$
\emph{does} look like a generalization of a familiar set thereof in
$\mathbb{C}^3$, which might raise our hopes.) Both in $\mathbb{R}^3$
and in $\mathbb{C}^3$, we can construct a maximal set of equiangular
lines by starting with a fairly nice fiducial vector and applying a
straightforward set of transformations. Is the analogous statement
true in $\mathbb{O}^3$?

An equiangular set of 27 lines would provide a map from the set of
density matrices for an ``octonionic qutrit'' to the probability
simplex in $\mathbb{R}^{27}$, yielding a convex body that would be a
higher-dimensional analogue of the Bloch ball. The extreme points of
this Bloch body, the images of the ``pure states'', might form an
interesting variety.

The 27 we have quoted here is related to a 27 that we encountered
above.  The algebra of self-adjoint operators on $\mathbb{O}^3$ ---
the ``observables'' for an octonionic qutrit --- is known as the
\emph{octonionic Albert algebra,} and it is 27-dimensional.  The group
of linear isomorphisms of~$\mathbb{O}^3$ that preserve the determinant
in the octonionic Albert algebra is a noncompact real form
of~$\mathrm{E}_6$~\cite{Baez:2002}.  As we saw earlier, the weights of
the minimal representation of the Lie algebra $\mathfrak{e}_6$ yield
the polytope $2_{21}$, from which we can derive the Hesse SIC
in~$\mathbb{C}^3$. An exact solution for an ``octonionic qutrit SIC''
might close this circuit of ideas.

It is possible to fit the Leech lattice into the traceless part of the
octonionic Albert algebra~\cite{Baez:2014}. This means that each point
in the Leech lattice is a ``Hamiltonian'' for a three-level octonionic
quantum system. This sounds a bit like a toy version of a vertex
operator algebra construction.

Having reached a point where the tone has taken a rather speculative
turn, we now embrace that attitude, just for the fun of it.

Another ``28'' that appears in the study of exceptional or unusual
mathematical objects is the size of the \emph{28-element Dedekind
  lattice.} This is a lattice in the sense of order theory, a
partially ordered set with the property that we can trace subsets of
elements upward through the ordering to where they join and downward
to where they meet. It is the \emph{free modular lattice on three
  generators} with the top and bottom elements removed, and Dedekind
showed how to construct it as a sublattice within the lattice of
subspaces of $\mathbb{R}^8$. Baez has suggested that its size is
therefore related to the Lie group $SO(8)$, which is
28-dimensional~\cite{Baez:2016b}. Without resolving this conjecture,
we note that the structure does sound a bit like something one would
see in quantum theory, or a close relative of it. The ``quantum
logic'' people have argued for a good long while that the lattice of
closed subspaces of a Hilbert space, ordered by inclusion, can be
thought of as a lattice of propositions pertaining to a quantum
system. If the system in question is a set of three qubits, then we'd
be talking about the lattice of subspaces of $\mathbb{C}^8$. To make
this look exactly like the setting of Dedekind's lattice, we would
have to do quantum mechanics over real vector space (``rebits''
instead of qubits), but that's not so bad as far as pure math is
concerned~\cite{Hardy:2012, Wootters:2013, Wootters:2013b}.

There's an idea, going back to Birkhoff and von Neumann in the 1930s,
that in quantum physics, we should relax the distributive law of
logic.  The argument goes that we can measure the position of a
particle, say, \emph{or} we can measure its momentum, but per the
uncertainty principle, we cannot precisely measure its position
\emph{and} its momentum at once.  Thus, we should reconsider how the
logical connectives
\begin{equation}
  \land = \hbox{``and''},\ \lor = \hbox{``or''}
\end{equation}
interact.  I'm not convinced this is really the way to dig deep into
the quantum mysteries: We can always impose an ``uncertainty
principle'' on top of a classical theory~\cite{Spekkens:2007,
  Spekkens:2016}.  Still, the idea is good enough to wring some
mathematics out of. In Boolean logic, we can distribute ``or'' over
``and'':
\begin{equation}
  a \lor (b \land c) = (a \lor b) \land (a \lor c).
\end{equation}
But if $a$, $b$ and $c$ are propositions about the outcomes of
experiments upon a quantum system, then we cannot combine them
arbitrarily and still have the result be physically meaningful.
``Complementary'' actions are mutually exclusive.  We should only
require that the distribution trick above works in restricted
circumstances.  When we organize all the propositions pertaining to a
quantum system into a lattice, we say that $a \leq b$ when $a$ implies
$b$.  Then, unlike Boolean logic, we say that the distributive trick
works when $a \leq b$ or $a \leq c$.  This makes our lattice
\emph{modular} instead of \emph{distributive.}

The free modular lattice on 3 generators is the structure we get when
we introduce a set of elements $\{a,b,c\}$ and build up by
combining them, assuming that the only restrictions are those due to
requiring that the lattice be modular.  In other words, it is the
logic we build by starting with three propositions and saying nothing
about them except that they should be ``quantum'' propositions, in
this very stripped-down way of being ``quantum''.

We can also try approaching from the $SO(8)$ direction.  Manogue and
Schray point out that we can label a set of 28 generators for $SO(8)$
in the following way~\cite{Manogue:1993}. 7 of them correspond to the
unit imaginary octonions $e_1$ through $e_7$. Then, for each of the
$e_i$, there are three pairs of unit imaginary octonions $(e_j, e_k)$
that multiply to $e_i$. These pairs give the other 21
generators. Considering the Fano plane, we have four generators for
each point: one for the point itself, and one for each line through
that point.

We can associate each generator with a line in $\mathbb{R}^7$ in the
following way. First, we label each point in the Fano plane by the
lines which meet there. For example,
\begin{equation}
(1, 1, 1, 0, 0, 0, 0)
\end{equation}
stands for the point at which the first three lines coincide. There
are seven such vectors, any two of which coincide at a single entry
(because any two points in the Fano plane have exactly one line
between them), and so any two of these vectors have the same inner
product. Next, we pick one of the three lines through our chosen
point, and we mark it by flipping the sign of that entry. For example,
\begin{equation}
(1, -1, 1, 0, 0, 0, 0).
\end{equation}
There are three ways to do this for each point in the Fano plane, and
so we get 28 vectors in all. Note that the magnitude of the inner
product is constant for all pairs. If the vectors correspond to
different Fano points, then their supports overlap at only a single
entry, and so the inner product is $\pm 1$. If they correspond to the
same Fano point, then the magnitude of the inner product is $1 - 1 + 1
= 1$ again.

So, the 28 of $SO(8)$ seems to be tied in with the other 28's: The
same procedure counts generators of the group and equiangular lines in
$\mathbb{R}^7$. (Counting --- a kind of math I understand, some of the
time.)

We now recall that the stabilizer subgroup for each vector in a
Hoggar-type SIC is isomorphic to the projective special unitary group
of $3\times3$ matrices over the finite field of order 9, known for
short as $\mathrm{PSU}(3,3)$.  Finite-group theorists also refer to
this structure as $\mathrm{U}_3(3)$, and as $\mathrm{G}_2(2)'$, since
it is isomorphic to the commutator subgroup of the automorphism group
of the \emph{octavians,} the integer octonions.  The symbol
$\mathrm{G}_2(2)'$ arises because the automorphism group of the
octonions is called $\mathrm{G}_2$, when we focus on the octavians we
add a 2 in parentheses, and when we form the subgroup of all the
commutators, we affix a prime.

We also recall that, given one SIC of Hoggar type, we can construct
another by antiunitary conjugation, and each vector in the first SIC
will be orthogonal to exactly 28 vectors out of the 64 in the other
SIC.  Furthermore, the Hilbert--Schmidt inner products that are not
zero are all equal.  Said another way, if we use the first SIC to
define a probabilistic representation of three-qubit state space, then
each vector in the second SIC is a probability distribution that is
uniform across its support.  Up to normalization, such a probability
distribution is a binary string composed of 28 zeros and 36 ones.

Let $\{\Pi_j^+\}$ be a SIC of Hoggar type, and let $\{\Pi_j^-\}$ be
its conjugate SIC.  Suppose that $U$ is a unitary that permutes the
$\{\Pi_j^+\}$.  Then a linear combination of 36 equally weighted
projectors drawn from $\{\Pi_j^+\}$ will be sent to a linear
combination of 36 equally weighted projectors from the set
$\{\Pi_j^+\}$, possibly a different combination.  But the only
sequences of 36 ones and 28 zeros that correspond to valid quantum
states are the representations of $\{\Pi_j^-\}$.  Therefore, a unitary
that shuffles the $+$ SIC will also shuffle the $-$ SIC.  Furthermore,
a unitary that \emph{stabilizes} a projector, say $\Pi_0^-$, must
permute the $\{\Pi_j^+\}$ in such a way that 1's go to 1's and 0's go
to 0's.

To repeat: Because each SIC provides a basis, we can uniquely specify
a vector in one SIC by listing the vectors in the other SIC with which
it has nonzero overlap.

A unitary symmetry of one SIC set corresponds to a permutation of the
other.  Using the second SIC to define a representation of the state
space, each vector in the first SIC is essentially a binary string,
and sending one vector to another permutes the 1's and 0's.  In
particular, a unitary that stabilizes a vector in one SIC must permute
the vectors of the second SIC in such a way that the list of 1's and 0's
remains the same.  The 1's can be permuted among themselves, and so can
the 0's, but the binary sequence as a whole does not change.

It would be nice to have a way of visualizing this with a more
tangible structure than eight-dimensional complex Hilbert
space --- something like a graph.  Thinking about the permutations of 36
vectors, we imagine a graph on 36 vertices, and we try to draw it in
such a way that its group of symmetries is isomorphic to the
stabilizer group of a Hoggar fiducial.  Can this be done?  Well,
almost --- that is, up to a ``factor of two'': It's called the
$\mathrm{U}_3(3)$ graph, and its automorphism group has the stabilizer
group of a Hoggar fiducial as an index-2 subgroup.

Now, is there a way to illustrate the structure of both SICs together
as a graph?  We want to record the fact that each vector in one SIC is
nonorthogonal to exactly 36 vectors of the other, that the stabilizer
of each vector is $\mathrm{PSU}(3,3)$, and that a stabilizer unitary
shuffles the nonorthogonal set within itself.  So, we start with one
vertex to represent a fiducial vector, then we add 63 more vertices to
stand for the other vectors in the first SIC, and then we add 36
vertices to represent the vectors in the second SIC that are
nonorthogonal to the fiducial of the first.  We'd like to connect the
vertices in such a way that the stabilizer of any vertex is isomorphic
to $\mathrm{PSU}(3,3)$.  In fact, because any vector in either SIC can
be identified by the list of the 36 nonorthogonal vectors in the
other, the graph should look locally like the $\mathrm{U}_3(3)$ graph
everywhere!  Is this possible?  Yes! The result is the
\emph{Hall--Janko graph,} whose automorphism group has the Hall--Janko
finite simple group as an index-2 subgroup.

The Hall--Janko group can also be constructed as the symmetries of the
Leech lattice cast into a quaternionic form~\cite{Tits:1980,
  Cohen:1980}.  Speculating only a little wildly, we can contemplate a
possible path from the Hoggar-type SICs in~$\mathbb{C}^8$ to the
Hall--Janko group and thence to the Leech lattice and the ``real SIC''
in~$\mathbb{R}^{23}$.

That's what we get when we think about the permutations of the 1's.
What about $\mathrm{PSU}(3,3)$ acting to permute the 0's?

This seems to lead us in the direction of the \emph{Rudvalis group.}
Wilson's textbook \booktitle{The Finite Simple Groups} has this to say
(\S 5.9.3):
\begin{quotation}
\noindent The Rudvalis group has order $145\ 926\ 144\ 000 =
2^{14}.3^3.5^3.7.13.29$, and its smallest representations have degree
28. These are actually representations of the double cover $2 \cdot
{\rm Ru}$ over the complex numbers, or over any field of odd
characteristic containing a square root of~$-1$, but they also give
rise to representations of the simple group ${\rm Ru}$ over the field
$\mathbb{F}_2$ of order 2.
\end{quotation}
Wilson then describes in some detail the 28-dimensional complex
representations of $2 \cdot {\rm Ru}$, using a basis in which $2^6
\cdot \mathrm{G}_2(2)$ appears as the monomial
subgroup~\cite{Wilson:2009}. And the 28 appears to be the
same 28 we saw before; that is, it's the 28 pairs of cube roots of the
identity in the octavians mod 2, so it's a set of 28 that is
naturally shuffled by $G_2(2)'$~\cite{Duncan:2010}.

The Rudvalis group can be constructed as a rank-3 permutation group
acting on 4060 points, where the stabilizer of a point is the Ree
group $^2 F_4(2)$. This group is in turn given by the symmetries of a
``generalized octagon''~\cite{Wilson:2010}. (Generalized $n$-gons
abstract the properties of the more familiar $n$-gons that, as graphs,
they have diameter $n$ and their shortest cycles have length $2n$. The
Fano plane is a generalized triangle.)  Compare this situation to the
Hall--Janko group, which has a rank-3 permutation representation on
100 points where the point stabilizer is $\mathrm{U}_3(3)$, and
$\mathrm{U}_3(3)$ is furnished by the symmetries of a generalized
hexagon~\cite{Schroth:1999}.

${\rm Ru}$ has a maximal subgroup given by a semidirect product
$2^6:\mathrm{G}_2(2)':2$.  This is what first caught my eye.
Neglecting the issue of the group actions required to define the
semidirect products, consider the factors: we have $\mathrm{G}_2(2)'$
from the Hoggar stabilizer, $2$ from conjugation and $2^6$ from the
three-qubit Pauli group.

This is reminiscent of a theorem proved by O'Nan~\cite{ONan:1978}:
\begin{quotation}
\noindent Let $G$ be a finite simple group having an elementary
abelian subgroup $E$ of order 64 such that $E$ is a Sylow 2-subgroup
of the centralizer of~$E$ in~$G$ and the quotient of the normalizer
of~$E$ in~$G$ by the centralizer is isomorphic to the group
$\mathrm{G}_2(2)$ or its commutator subgroup $\mathrm{G}_2(2)'$.  Then
$G$ is isomorphic to the Rudvalis group.
\end{quotation}

The peculiar thing is that the Hall--Janko group is part of the
``Happy Family'', i.e., it is a subgroup of the Monster, while the
Rudvalis group is a ``pariah'', floating off to the side.  The Hoggar
SIC almost seems to be acting as an intermediary between the two
finite simple groups, one of which fits within the Monster while the
other does not.

Finally, what connects the largest of sporadic simple groups with the
second-smallest among quantum systems?

I was greatly amused to find the finite affine plane on nine points
also appearing in the theory of the Monster group and the Moonshine
module~\cite{Miyamoto:1995, Gannon:2004}.  In that case, the 9 points
and 12 lines correspond to involution automorphisms.  All the
point-involutions commute with one another, and all the
line-involutions commute with each other as well.  The order of the
product of a line-involution and a point-involution depends on whether
the line and the point are incident or not.

This is probably of no great consequence --- just an accident of the
same small structures appearing in different places, because there are
only so many small structures to go around.  But it's a cute accident
all the same.

\bigskip

I thank Marcus Appleby and Markus Grassl for correspondence, and John
Baez for the opportunity to discuss the sporadic SICs and their
conceptual fellow travelers at the $n$-Category Caf\'e. This research
was supported in part by the John Templeton Foundation. The opinions
expressed in this publication are those of the author and do not
necessarily reflect the views of the John Templeton Foundation.

\bibliographystyle{utphys}

\bibliography{exceptional}

\end{document}